# Phase transition or Maxwell's demon in Granular gas?

## P. Jean, H. Bellenger, P. Burban, L. Ponson
##    P. Evesque

**Lab MSSMat,  UMR 8579 CNRS, Ecole Centrale Paris**
**92295 CHATENAY-MALABRY, France,** e-mail: evesque@mssmat.ecp.fr

**Abstract:**

*Dynamics of vibro-fluidised granular gas is investigated experimentally using the transfer of grains from a compartment through a horizontal slit at a given height* h *. It is demonstrated that the transfer rate* j *varies linearly with the grain number* N *in the box when* N *remains small; however* j(N) *becomes strongly non linear as soon as the number* n *of layers is larger than 0.3;* dj/dN *becomes negative for* n>0.4. *It is found also that the maximum of* j(N) *increases slightly with the acceleration* $a\omega^2$ *of the vibration which excites the granular gas. Using dynamical system theory, dynamics equations are written, and a critical bifurcation is found,  which explains the existence of a condensation and of a phase transition. This explains how the pseudo " Maxwell's demon" works in granular gases. This experiments contradicts recent modelling .*

**Pacs # : 5.40 ; 45.70 ; 62.20 ; 83.70.Fn**

Recently, granular matter has been the topics of numerous works; this is probably because it has been taken as an archetype for studying complex dynamical systems, and to illustrate many non linear effects. Among these effects, one of them, which is known as Maxwell's demon, looks quite impressive: when shaking 2 half-boxes connected via a slit and partly filled with granular matter one observes under some special excitation conditions the partial emptying of one cell into the other one [1, 2]. This phenomenon can be observed with dense layers [1] or in the case of granular gas [3]. It is not the purpose of this paper to discuss the correctness of the terminology "Maxwell's demon" to explain this behaviour, neither to discuss the applicability of this notion to classic thermodynamics problem [4].

    This paper reports on a "toy" experiment dedicated to initiate undergraduate students, *i.e.* P. Jean, H. Bellenger, P. Burban, L. Ponson, to the physics (i) of granular media, (ii) of dynamical system theory and of chaos and (iii) of phase transition. The experiment is quite simple and can be home made. Of course, it can be improved quite a bit; but as it is, it allows already to illustrate few important concepts and to point out also few incompatibilities and mismatches between experiments and recent theories published by famous journals   [3, 5]. In particular it points out the difficulty of applying to granular gas a continuous formalism in general, and the fluid mechanics approach, as it has been proposed in [5].

    Furthermore, this experiment shows that the transfer rate $j_{1\rightarrow 2}$ from one compartment to the other one varies linearly with the grain number N, at small N; this contradicts the theory proposed in [3], which finds that j(N) should scale as $N^2$. It

shows also that $j_{1\to 2}$ passes through a maximum when N is increased; then $j_{1\to 2}$ decreases with increasing N at larger N. It results in an unstable behaviour and one compartment fills up spontaneously while the other one empties.

We have found that $j_{1\to 2}$ starts behaving non linearly with N as soon as the number of layer n in one cell is larger than 0.2 (and the maximum of j occurs at 0.3-0.4 n); this proves that energy loss due to grain-grain collision modifies deeply the mechanical behaviour of a granular gas as soon as particles can collide; this is in agreement with recent interpretation [6] of micro-gravity experimental results [7] on clustering of vibro-fluidised gas. Indeed these experiments prove an other time that homogeneous granular gas only exists in the Knudsen regime, that is to say when particles do not touch one another because their mean free path $l_c$ is larger than the box length L. This experiment shows also that clustering occurs as soon as particles collide and as soon as their mean free path $l_c$ becomes smaller than the box length L.

Indeed this is a new finding: of course clustering of granular gas has been predicted already [3,5,8,9]; but these works use a continuous approach which is not valid when $l_c<L$, so that they do not apply to granular clustering and their predictions cannot be considered as correct. This incompatibility will be confirmed by studying the flow j(N) *vs*. N of particles from a compartment which contains a small number N of beads. We will find that j(N) scales as N in the present work; but it is predicted to scale as N² in [3], which disclaims this paper, hence the continuous approach it uses. On the contrary the present experiment and the clustering experiment in weightlessness conditions [6-7] both demonstrate that a more general approach is required, which will not be based on the hypotheses of a continuum medium. But this is a much harder task!

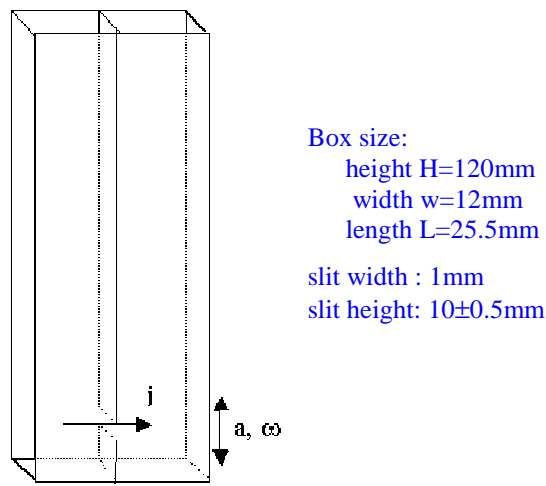

Box size:
  height H=120mm
  width w=12mm
  length L=25.5mm

slit width : 1mm
slit height: 10±0.5mm

*Fig. 1: Sketch of the box :*

## 1. Experimental set-up:

A 3d rectangular box (height H=120mm, width w=12mm and length L=25.5mm) is divided into two equal parts by a vertical wall of thickness δe=1mm. These two parts are connected via a horizontal slit (width 1.5mm) located in between $h_1$=9mm and $h_2$=10.5mm from the bottom. This box contains from 0 to 3000 bronze spheres from Makin Metal Powders; their diameter d ranges in between d=0.425mm and d=0.600mm and their average mass is 0.72-0.76 mg, so that a dense layer contains (L-δe)w/(2d²sin60)≈530-550 beads in mean, in each half box. The slit allows the grains to pass from one part of the box to the other one when the box is vibrated at large enough amplitude, as sketched in Fig. 1.

So, the box is mounted on a loudspeaker which vibrates vertically (amplitude a, frequency $f=\omega/(2\pi)$). The system exhibits a mechanical resonance around 18-20 Hz, which is also the working frequency range during the present experiment. The loudspeaker is excited via an ac generator and an amplifier. Amplitude a of vibration has been measured optically; voltage, frequency and duration of loudspeaker excitation has been measured using a numerical Agilent 54621A oscilloscope. Grain number in each compartments and grain flow have been determined either by weighting or by direct counting.

## 2. Experimental results :

### 2.1. *Distribution at equilibrium as a function of the acceleration* Γ

Be $N_1$ and $N_2$ the number of beads in each compartment. In a first experiment, the number $N_{tot} = 2N_o = N_1 + N_2$ of beads has been fixed to 460 and the acceleration has been changed by varying the voltage applied to the loudspeaker at constant frequency *f*=19.5 Hz. At large amplitude of vibration, the balls are equally distributed into the two half boxes. Lowering the acceleration Γ provokes the asymmetry of the distribution below a given $\Gamma_c$: one of the box is more filled than the other one, the smaller Γ the larger the asymmetry and the larger $|\Gamma_c - \Gamma|$ the larger the asymmetry. One can measure the asymmetry through the parameter $\varepsilon = N_1/N_{tot} - 1/2$.

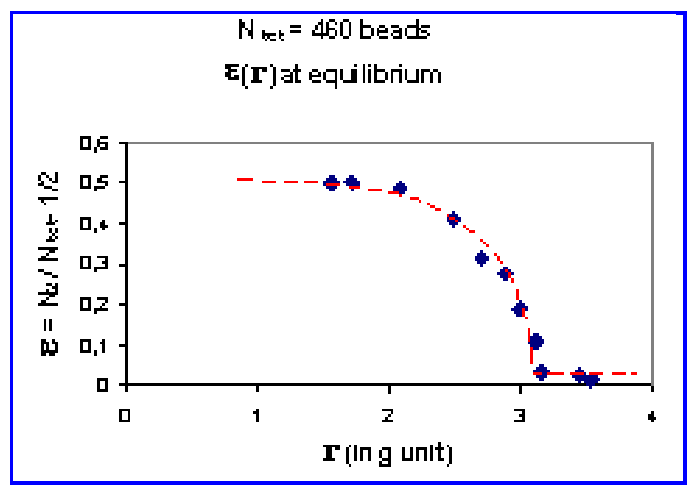

*Figure 2: Variation of the asymmetry of the equilibrium distribution of beads as a function of the acceleration Γ in g unit, when the total number $N_{tot}$ of beads in the two compartments is $N_{tot}$ =460. The asymmetry is measured via $\varepsilon(N_{tot}) = N_1/N_{tot} - 1/2 = 1/2 - N_2/N_{tot}$.*

One can then analyse this phenomenon in terms of the bifurcation theory. In this case, Γ plays the part of the control parameter and $\varepsilon = N_1/N_{tot} - 1/2$ the part of the order parameter; their interdependence is displayed in Fig. 2, for $N_{tot}$=460 beads. As shown in this Fig. 2, the evolution of ε is continuous, but the slope ε *vs.* Γ is discontinuous and changes from vertical to horizontal when Γ→$Γ_c$ ; so this bifurcation is critical.

One observes also in Fig. 2 that ε is slightly different from 0 when acceleration Γ is above $Γ_c$. However, this difference is never large and does not overpass the typical amplitude of fluctuations $δε=δN/N_{tot} =1/\sqrt{N_{tot}}$ one shall expect in application of the Gaussean central limit theorem; this is normal and expected due to the small number of beads involved in this experiment.

## *2.2. Distribution at equilibrium as a function of $N_{tot}$ :*

We demonstrate in the following that these results, *i.e.* both the threshold $Γ_c$ and the curve ε(Γ), depend on the bead number $N_{tot}=2N_o$ . For instance, one can repeat the experiment of §-2.1 but one can vary $N_{tot}$ instead of a, and let a, ω and Γ fixed; one obtains ε($N_{tot}$); a typical example is given in Fig. 3 which reports ε($N_{tot}$) curve at Γ=2.6g and f=18.5 Hz. In this case, one observes an other "critical" bifurcation : ε remains constant and vanishingly small as far as $N_{tot}$< 350; then it starts growing fast and saturates at ε=0.5 when $N_{tot}$>700. Indeed, ε=0.5 means that one of the compartment is merely empty and that the other one contains most of the grains.

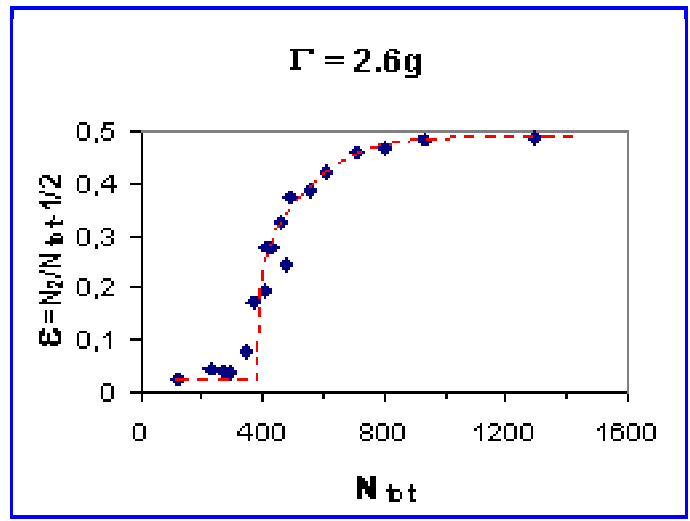

*Figure 3: Variation of the asymmetry ε($N_{tot}$) of the equilibrium distribution of beads at a given acceleration Γ= 2.6 g, as a function of the total number of beads $N_{tot}$ in both compartments .The asymmetry is measured via ε($N_{tot}$) =$N_1/N_{tot}$-1/2=1/2-$N_2/N_{tot}$.*

The curve ε($N_{tot}$) exhibits some noise. Another time, as in §-2.1, the amplitude of this noise is compatible with normal fluctuations due to the small number of grains in each compartment in the vicinity of ε=0 or ε=0.5. However, the transition around $N_{tot}$=300-450 in Fig. 3 looks rather smooth, and noisier; at least, it does not look as sharp as the one occurring at Γ=3.2g in Fig. 2. Does it means it does not correspond to a critical bifurcation in this representation space? We believe not; we think this broadening is linked to finite size effect and to critical fluctuations;  but we will discuss this point later.

## *2.3. Kinematics to equilibrium:*

The kinematics of reaching the equilibrium has been investigated in the condition of Fig. 2, *i.e.* for $N_{tot}=460$ and for three different acceleration. Results are reported in Fig. 4. The study starts with equi-repartition; and the evolution of ε as a function of time has been recorded by stopping quickly the vibration at different instants after the beginning; then the amount of beads in each compartment has been weighted and placed back into the compartments. And vibration has been applied for a new lapse of time, and so on… The quick stoppage of the vibration is equivalent to a fast freezing of the system.

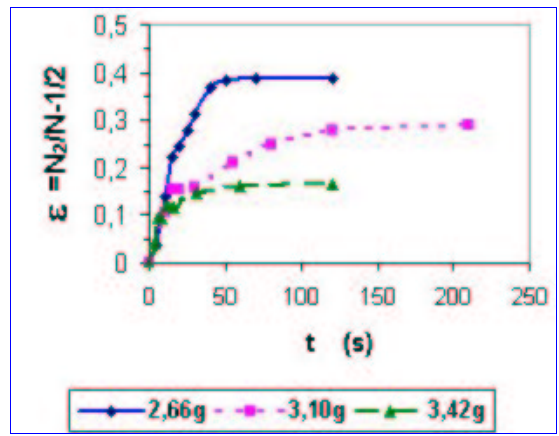

***Figure 4:*** *Time dependence of the distribution asymmetry for a total bead number $N_{tot}=460$, at different accelerations* Γ*, (*Γ*=2.7g, 3.1g and 3.4g). The asymmetry is measured via* $\varepsilon(N_{tot}) = N_1/N_{tot}-1/2=1/2-N_2/N_{tot}$.

Indeed the evolution of the distribution occurs because the flows $j_{1\to2}$ & $j_{2\to1}$ of particles through the slit are different. Equilibrium occurs when the two flows are equal. So one may describe the evolution with the set of coupled equations:

$$dN_1/dt = -j_{1\to2}+ j_{2\to1} \quad \& \quad dN_2/dt = j_{1\to2} - j_{2\to1} = -dN_1/dt \qquad (1)$$

In general, each current j can depend on the number of beads in both compartments and on the density of grains in the slit. In the present case however, as both compartments are not dense and as the current of grains flowing through the slit seems to be small, one may suppose that each current $j_{1\to2}$ (*resp.* $j_{2\to1}$) depends only on the conditions at work in the compartment from which the grains flow, *i.e.* compartment 1 (*resp.* compartment 2) for $j_{1\to2}$ (*resp.* $j_{2\to1}$).

Owing to Eq. (1) one can conclude that measuring the evolution of ε or of $N_1-N_2$ does not allow to determine $j_{1\to2}$ and $j_{2\to1}$, but only their difference $j_{1\to2} - j_{2\to1}$. However, one way to proceed to determine unambiguously $j_{1\to2}$ is to force compartment 2 to remain empty and to record the flow from compartment 1 as a function of time (and hence on $N_1$). This is what has been done.

## 2.4. Flow $j_{1\to 2}$ of beads through the slit :

So, in order to get a better insight in the physics of this dynamical system, one can decide to study the kinematics of a single half box. This can be simply performed using some sellotape in the second box. This sellotape acts as a glue which traps the grains as they are transferred from the first compartment. So starting with a large number of grains N in the first compartment and stopping vibrating intermittently to count and eliminate the number of glued grains in compartment 2 allows to determine $N_1(t)$ and its time derivative $dN_1/dt$ which is the flow $j_{1\to 2}=dN/dt$. Indeed, one expects this flow to depend on the vibration parameters and on the number N of grains only; so this allows to determine $j_{1\to 2}(N)$. This process can be repeated for different values of the acceleration Γ. This study is reported in Figs. 5 & 6 for 3 different accelerations. We report also in Fig. 6 the variations of j as a function of the number n of layer of beads. In the present case, the number of beads per layer is 540 about.

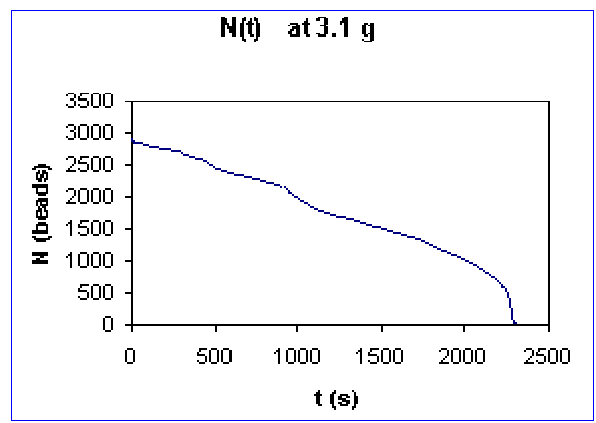

*Figure 5:* *Typical time dependence of the number* N(t) *of grains in the first compartment, keeping empty the second one, at different accelerations* Γ, Γ=2.7g, 3.1g *and* 3.4g.

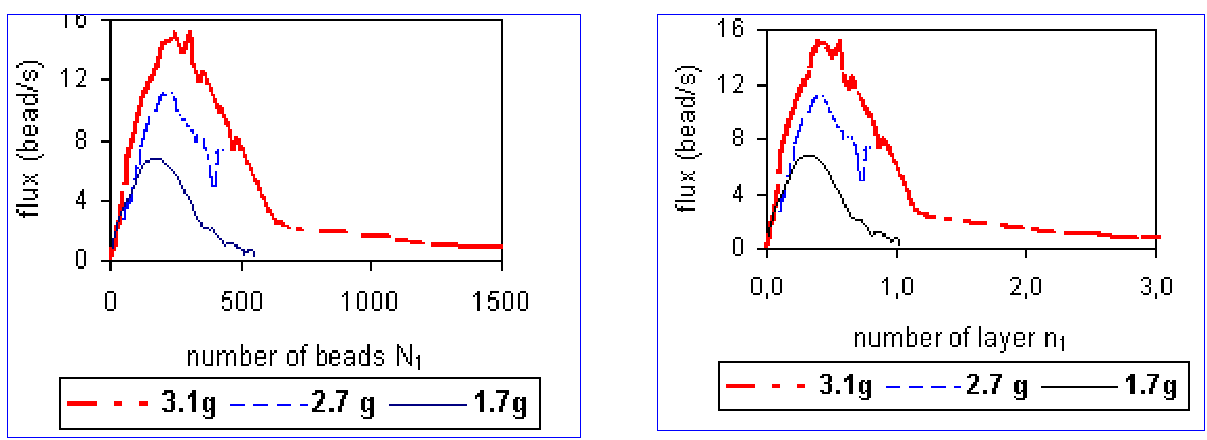

*Figure 6:* *Flow* $j_{1\to 2}$=dN(t)/dt *of grains flowing from the first compartment as a function of* N *of grains in compartment #1(left Fig.) or as a function of the number* n *of layers in compartment #1 (right Fig.), keeping empty the second compartment, at different accelerations* Γ, Γ=2.7g, 3.1g *and* 3.4g. *It is worth noting that (i)* j *varies lineally with* N *(or* n*) at small N, (ii) that* j *exhibits a maximum, which occurs much before* n= ½ *, and (iii) that non linear behaviour occurs at* n ≥ ¼ *.*

It is worth noting that a precise evaluation of j requires to take rather large increment of time since the normal expected fluctuations on $j\Delta t=\Delta N$ is $\sqrt{\Delta N}$. In the present case, the time increment which has been used for Fig. 6 has been 10s about; so the normal fluctuations depends on j; it lays around $\Delta j/j=10\%$ when j=10 grains/s and $\Delta j/j=30\%$ when j=1 grains/s.

We note in Fig. 6 that j varies linearly with N (or n) at small N, *i.e.* j=AN. This is in contradiction with the scaling found theoretically by Eggers [3], who gives $j_{1\to 2}=F_{l\to r}=F_o\,N_l^2\,\exp(-aN_l^2)$. This cast then a serious doubt on the validity of Eggers' theory.

We note also in Fig. 6 that j passes through a maximum $j_m$ located at $N_m$ (or $n_m$); both $j_m$ at $N_m$ depend on $\Gamma$; and the larger $\Gamma$, the larger $j_m$ and $N_m$. We note also that $N_m$ varies from 190 to 250 grains in the present case, depending on $\Gamma$ ($\Gamma$ ranges from 1.7 to 3.2 g). As it will be explained in §-2.6, a linear variation of $N_m$ with $\Gamma$ explains the existence of a critical bifurcation when lowering $\Gamma$ at fixed $N_{tot}$. However, this requires first to understand the effect of a maximum in the j(N) curve, which is explained in §-2.6.

It seems also that the slope A of the curve j *vs.* N is rather constant near the origin N=0, and that this slope A is rather independent of $\Gamma$. This may be an artefact due to the small number of grains involved since the phenomenon is observed when N<20-50 grains; indeed, N<50 limits strongly the accuracy on (i) the number of grains and (ii) on the flow, and leads to strong uncertainty. A best way to estimate the slope consists then in analysing the time dependence of the data at small N and to fit with an exponential law, since from Eq. (1) one expects the emptying law to be:

dN/dt=-AN  (2.a)

whose solution is

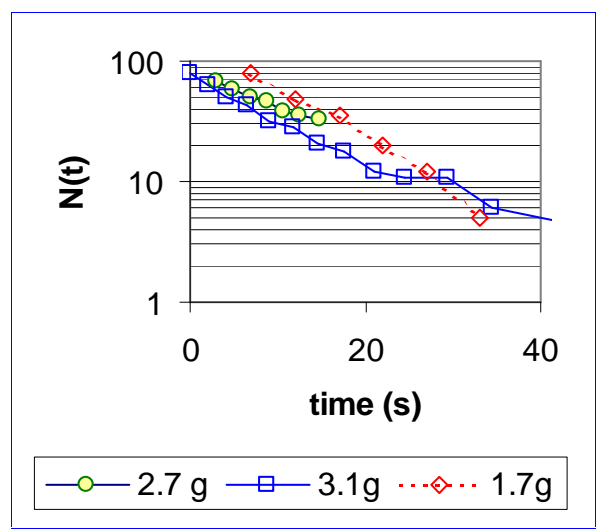

*Figure 7: Time dependence of the emptying when N is small, in semi-log plot, for 3 different accelerations $\Gamma$. The time origin of the three curves have been translated from an arbitrary value. The 3 curves exhibit the same time dependence.*

$$N = N_o \exp(-At) \qquad (2.b)$$

This fit is tested in Fig. 7. It works rather well, and gives about the same lifetime $\tau=1/A \approx 16s$ for the three accelerations; so it confirms the result of Fig. 6.

## *2.5. Validation of the hypotheses of Eq. (1):*

We see also from Fig. 6 that the flow of beads through the slit is always small (<10 beads/s). But the typical height the beads occupy in their gaseous mode is limited to 1cm about due to gravity; this leads to a typical speed $v_t=(2gh)^{1/2} \approx 0.4$m/s and to a typical time $\tau_1=(2h/g)^{1/2} \approx 0.04$s . So the transfer of a bead from a compartment to the other one shall not last more than 0.05s ; then the probability that a collision between two beads occurs when a bead passes through the slit is as little as : $p < j_{max}\tau_1 d^2/[(w-d)(\delta e-d)]=0.01$ . This makes the risk of bead-bead collision very little and quite negligible, which validates the hypothesis made earlier. In the same way, the occupancy of each compartment is small so that the transfer rate from box 1 to box 2 shall not depend on the occupancy of box 2. This validates then completely the model of transfer rate used in Eq. (1).

It is worth noting however that these assumptions are not always valid when observing Maxwell's demon in granular media. For instance, it is not true in the case of experiments reported in [1] with deep layers of sand; in this case, collisions which occur in the slit are frequent and the probability of finding a hole in compartment 2 which is large enough to accept a grain coming from the other compartment is rare; this makes the mechanics of the dense system completely different, even if one feels at first sight that it is the same demon which is at works in both cases.

## *2.6. Use of Fig. 6 to understand (or predict) clustering:*

We demonstrate now that Figs. 2 and 3 can be deduced from Fig. 6:

First of all, we remark that the curve j(N) exhibits a maximum $j_{max}$ at $N_m$. We note also that both $j_{max}$ and $N_m$ depend on $\Gamma$ (*cf.* Fig. 6), and the larger $\Gamma$ the larger $j_{max}$ and the larger $N_m$ (and $n_m$).

Let us first consider the case when both compartments contain less beads than $N_m$; in this case, the solution $N_1=N_2$ is stable against any small perturbation, since if $N_1>N_2$ implies $j(N_1)>j(N_2)$. So $N_1=N_2$ is an attractor of the dynamics.

Conversely, $N_1=N_2$ becomes no more an attractor of the dynamics when $N_1>N_m$ & $N_2>N_m$ since $j(N_1)>j(N_2)$ if $N_1<N_2$ (which implies that the emptier compartment keeps on emptying and the fuller one keeps on filling up). This explains why the system looks for an other solution. Indeed, there are two new solutions when $N_{tot}>2 N_m$, which are symmetric compared to $N_m$ . They both satisfy the set of Eqs. (3):

$$N_1+N_2=\text{constant}= 2N_o>2N_m, \qquad (3.a)$$

$$j(N_1)=j(N_2) \qquad (3.b)$$

$$N_1<N_m \text{ and } N_2>N_m \qquad (3.c)$$

More precisely, the stability of these new solutions require $dj/dN_1+dj/dN_2>0$, as can be deduced from perturbation and their existence needs that the two different solutions $N_1(j)$ & $N_2(j)$ have a mean $(N_1(j)+N_2(j))/2$ which increases continuously when j decreases from $j_{max}$ (this will be shown in a next paper).

In the present case both conditions are satisfied indeed; this is why a critical bifurcation is found when $2N_o>2N_m$. Furthermore, using limited development in the vicinity of $2N_o=2N_m$ and writing $N_1+N_2=2\ N_o= 2N_m+\delta N$, one finds $N_1-N_2\approx\delta N^{½}=2\varepsilon$. This explains why the slope $\varepsilon(N)$ changes from horizontal to vertical when passing through $N=N_m$ keeping $\Gamma$ constant and increasing continuously N; this is why also $N=N_m$ is a bifurcation point at $\Gamma$=constant.

Furthermore, as pointed out already, $N_m$ and $j_{max}$ depend both on $\Gamma$; be $N_m=N_{mo}+\beta\Gamma$ the law of variation; it is a good approximation as shown in Fig. 8. So, according to this modelling, one finds that the solution $N_1=N_2=2N_o$, is stable when $\Gamma>\Gamma_c=(N_o-N_{mo})/\beta$ ; but that it becomes unstable when $\Gamma<\Gamma_c$ ; this generates an heterogeneous distribution as soon as $\Gamma<\Gamma_c$. Using limited development and the linear variation of $N_m$ *vs.* $\Gamma_c$ , one gets that $2\varepsilon=N_1-N_2$ shall scale as $(\Gamma_c-\Gamma)^{½}$ in the vicinity of $\Gamma_c^-$, *i.e.* when $\Gamma<\Gamma_c$. So j(N) given in Fig. 6 predicts that the slope of $\varepsilon(\Gamma)$ changes also from vertical to horizontal at $\Gamma=\Gamma_c$ as it is observed in Fig. 2. The bifurcation which occurs at $\Gamma=\Gamma_c$ is then also a critical bifurcation.

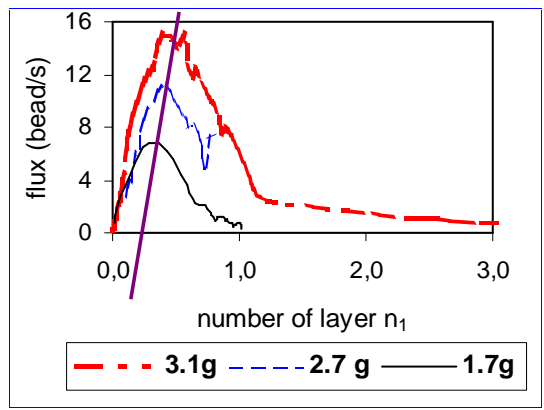

*Figure 8:* *Linear dependence of* $j_{max}$ *and* $N_m$ *with* $\Gamma$.

Things can be a bit more complicated, depending also on the values and of the signs of larger-order derivative $(d^p j/dN^p)_{N_m}$ of j(N) at $N_m$. In particular one can get in some case some sub-critical bifurcation… This will be studied in a next paper.

This next paper shall also discuss what are the right parameters which govern the flow: is it the typical acceleration $\Gamma=a\omega^2$ of the box, or its typical speed $v=a\omega$? Indeed, one can expect it is the same as the one which governs the physics of the gas of particles. So, as remarked in [10], one shall then expect it is **not** governed by the acceleration $\Gamma=a\omega^2$, but rather by the speed $v=a\omega$ of the vibration; otherwise, one gets misunderstanding of granular gas physics as in [11].

As in a granular gas, the particles do not interact strongly, their dynamics can be well represented by the one of a single bead in a vibrating box, whose statistics is described in [12] when gravity is strong enough to confined the granular gas.

However, when the box has a lid on top and when the vibration amplitude becomes large enough to allow the bouncing of the balls with this lid too, then gravity becomes negligible; in this case, the model of a single particle in a vibrating box without gravity [13] should apply. On the other hand, when the system contains few layers of particles, the dynamics is approximately the one of a cake which performs periodic free-flight followed by completely inelastic collisions with the vibrating plate; this looks like the problem of the bouncing of an inelastic ball which is treated in [14].

In order to show the difference which is caused by a lid, we have also investigated this case in order to prepare a flight experiment in weightlessness condition. This is reported in the next section. It gives also a good example of the efficiency of the model of the bouncing of an inelastic ball, when both the number of layers and the acceleration are large.

## *2.7. flow from a compartment closed on top:*

Investigation of larger vibration intensity requires the use of a box closed on top. This was achieved with a box having the same horizontal section, and the same location and geometry of slit; but the height of the cell is H=30mm. Results are reported in Fig. 9. The experiment has been performed with a mechanical vibrator, working at constant amplitude of vibration, *i.e.* a=1.7mm, but at different frequencies f, in the range 17Hz<f<33Hz, so that the accelerations range is $2g <\Gamma< 7g$. Data have been obtained by weighting the mass which has flown from the slit every 5 s.

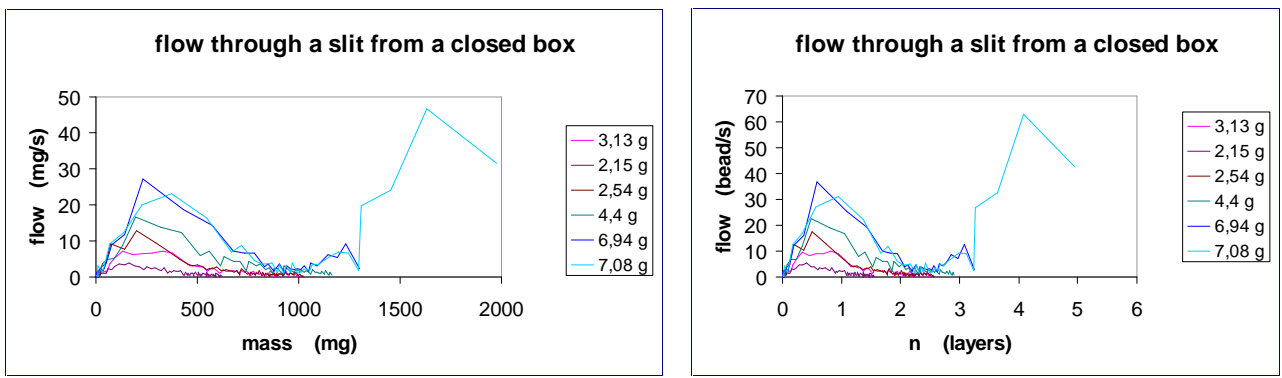

*Figure 9:* Flow $j_{1\rightarrow2}$=dN(t)/dt *of grains from the slit of a vibrated compartment as a function of the number N of grains in the compartment. vibration amplitude* a=1.7 mm*, variable frequency* f $=(\Gamma/a)^{1/2}$ /$(2\pi)$. *Left Fig.: N and j are given in mass and mass/s . Right Fig: N is given in number of layers* n*, and j in beads/s; bead mass=0.74mg ; there are 540 grains later. Differences with Fig. 6 experiment are: compartment is closed on top; frequency is varied and a fixed. One observe similar trends; one see also a second increase for large acceleration and large n. In this case, the motion looks like the one of an inelastic which perform free flight; when Γ is large enough, the cake can reach the slit and the beads flow out. This requires* h/a=6>Γ/g *about.*

The curves j(N) exhibit behaviours similar to the ones of Fig. 6. As in Fig. 6, and for the same reason, the curves fluctuate due to the small amount of beads involved. A better precision on the mean behaviour is possible to obtained by repeating the same experiment few times and by averaging. However this will not reduce the natural uncertainty on each experiment, because it is linked really to the small number of beads flowing from the slit.

As in Fig. 6, the slope of j(N) at N=0 seems to be independent of $\Gamma$; but the accuracy of the determination remains small; We have observed also that no flow passes through the slit if the number n of layer is too large and when the acceleration is small enough, i.e. when $\Gamma<\Gamma_c$; so there is a critical limit $n_c$ above which no flow occurs when $\Gamma<\Gamma_c$; indeed $n_c$ depends also slightly on $\Gamma$, the larger $\Gamma$, the larger $n_c$.

But this is no more true for larger $\Gamma$s; in this case one finds that the beads flow always from the slit; indeed, the system looks like a cake which makes inelastic bouncing on the vibrator in this case; so one expects that the height of free flight can reach the slit height h when $a^2\omega^2>gh$, which reads also $\Gamma>\Gamma_c=g\,h/a$, since $\Gamma=a\omega^2$. Indeed, one observes this regime in Fig. 9 for larger values of n and $\Gamma$.

## 3. Discussion and conclusion:

This simple experiment allows to illustrate efficiently a series of concepts which are required in non linear physics, dynamical system theory and phase transitions; in particular, it shows that it is better to measure the current j, rather than the balance of the transfer. It exemplifies also that the real motor of the instability comes the decrease of j with the control parameter, which is N in the present case. Indeed, this results in the instability of the transfer balance : from Eq. (1) one can write $d(N_1-N_2)/dt=-2\{j_1(N_1)-j_2(N_2)\}$ ; so and expanding j around its pseudo-equilibrium value $j_o$ at $N_o=(N_1+N_2)/2$. One gets:

$$d(N_1-N_2)/dt=-2A\,(N_1-N_2) \qquad (4)$$

where A is the slope of j at $j_o$. Eq. (4) shows then that population difference grows exponentially with time if A is negative, *i.e.* when the slope of j is negative, while it decreases exponentially if A is positive. And the characteristic time $\tau$ is $\tau=|1/A|$.

One gets then that $N_1=N_2$ is the stable equilibrium if A>0, but it is not when A is negative. Indeed, when A is negative, the system diverges till the difference $N_1-N_2$ becomes large enough to reach a new basin of attraction (when it exists); this new point is characterised by some A' and the system converges towards this new equilibrium, which is stable if A'>0.

This is a common feature of all instabilities. Also, when a state becomes unstable by the modification of a control parameter, this is because A changes from positive to negative. When A varies continuously, that means that A crosses 0 at the bifurcation and that the characteristic times becomes infinite. It means then that the dynamics slows down when A tends to 0. This is called the critical slowing down and the bifurcation is called critical bifurcation.

In general dynamical systems are a bit more complicated, because they are described by a set of few parameters, *i.e.* few dimensions; each of them having a different $A_i$. When all the $A_i$ are positive the point is an attractor and the point correspond to a stable equilibrium, otherwise it is not an attractor; however some of the direction can be attracting and some others repulsive...

It is also obvious that the clustering process which has been studied looks like a phase transition, since one observes the equilibrium between a dense "liquid" phase and a "gaseous" one when $N>N_m$. This connects the physics of phase transition to the one of bifurcation. However, a difference between these two problems exists because the interface between the liquid and the gas builds up spontaneously in the liquid-gas transition, while the wall and the slit forces the position of the interface in the present clustering. So these two transitions are not strictly equivalent.

Also, j is a current; so Eq. (1) is related to a problem of diffusion; this can be shown easily if one takes a larger number of interconnected boxes, either in 1d or in 2d. So, Generalisation of Eq. (1) can be performed easily and transformed into a Master equation, into a diffusion equation or into a Fokker-Plank equation,… This is then a simple way to initiate students to diffusion problems and to its different modelling, to approach the continuous limit. It will illustrate also efficiently the effect of a negative diffusion coefficient, which is a leading point when one is concerned with phase transition.

Some of the explanations which have been given in this paper can be developed a bit further and the notions can be improved. For instance one can try to predict more precisely the kind of bifurcation from the shape of j(N). Also, as the "granular gas" occurs only when the particles do not collide together, a good modelling seems to be the one of a bouncing ball; so one can then try to use this modelling, to get a better insight of the "granular gas" phase and to calculate some parameters. For instance, we will show in a forthcoming article that the mean free path of the grains in this "granular gas" phase depends only on the number of layers n (and not really on h or g). This will confirm that the gaseous regime occurs only in the Knudsen regime, when no continuous mechanic equation can be written...

As a conclusion, it turns out that this simple "toy" experiment allows to touch with the thumb some important problems of physics and to illustrate some theoretical concepts (physics of granular media, bifurcation theory, limits between continuous *vs.* discrete approach, diffusion theory). It can be also extrapolated in few different ways… Furthermore, it has already served to demonstrate that few hypotheses or conclusions published recently in famous papers [3,5] were not correct. This is indeed a good point to teach to students, because this is the only way that science can use : asking new questions, looking for the validity of the approximations used and of the hypotheses. And this does not require complicated arguments, but rather good thinking and clear mind. It shows also that scientific truth has always to be questioned, demonstrated and argued… and is not a question refereeing.

So it is really a powerful experiment.

*Acknowledgements:* CNES is thanked for partial funding.

*Present e-mail addresses of the student team*:
   jeanp3@cti.ecp.fr, ponsonl3@cti.ecp.fr, burbanp3@cti.ecp.fr, bellenh3@cti.ecp.fr

The electronic arXiv.org version of this paper has been settled during a stay at the Kavli Institute of Theoretical Physics of the University of California at Santa Barbara (KITP-UCSB), in june 2005, supported in part by the National Science Fundation under Grant n° PHY99-07949.

*Poudres & Grains* can be found at :
http://www.mssmat.ecp.fr/rubrique.php3?id_rubrique=402